# Extreme electron-photon interaction in perovskite glass


*Sergey S. Kharintsev[1]\*, Elina I. Battalova[1], Ivan A. Matchenya [2] Alexander A. Marunchenko[2] and Anatoly P. Pushkarev[2]*

[1]Department of Optics and Nanophotonics, Institute of Physics, Kazan Federal University, Kazan 420008, Russia

[2]School of Physics and Engineering, ITMO University, St. Petersburg 197101, Russia

skharint@gmail.com



ABSTRACT

The interaction of light with solids can be dramatically enhanced owing to electron-photon momentum matching. This mechanism is driven by either quantum confinement or long-range structural correlations in media with crystal-liquid duality. In this paper, we address a new strategy based on both phenomena for enhancement of the light-matter interaction in a direct bandgap semiconductor – lead halide perovskite $CsPbBr_3$ – by using electric pulse-driven structural disorder. The disordered (glassy) state allows the generation of confined photons, and the formation of an electronic continuum of static/dynamic defect states across the forbidden gap (Urbach bridge). Both mechanisms underlie photon-momentum-enabled *electronic Raman scattering* (ERS) and single-photon anti-Stokes photoluminescence (PL) under sub-band pump. PL/ERS blinking is discussed to be associated with thermal fluctuations of cross-linked $[PbBr_6]^{4-}$ octahedra. Time-delayed synchronization of PL/ERS blinking causes enhanced spontaneous emission at room temperature. Our findings indicate the role of photon momentum in enhanced light-matter interactions in disordered and nanostructured solids.




**INTRODUCTION**

Defects that inevitably occur in solids when natural growth or synthesis are mainly associated with static lattice imperfections (vacancies, impurities, interstitials, etc.) affecting their optical and electronic properties. In addition, thermal and polar fluctuations of a lattice can invoke dynamic disorder[1,2] which produces an electronic continuum of dynamic states across the forbidden gap, known as the Urbach bridge.[3] Physically, it leads to the hybridization of static and dynamic defect states,[4–6] allowing charge-carrier hopping either through inelastic light scattering or tunneling. This behavior is typical for solids possessing crystal-liquid duality[7] in which vibrations of coupled rigid frameworks are long-range correlated. Such systems include perovskites,[7–9] mxenes,[10,11] chalcogenides,[12] liquid crystals,[13] and high-entropy crystals.[14] Structural dynamics in their lattices can significantly influence the electronic polarizability, breaking the Franck-Condon approximation.[15] The static/dynamic states can be disentangled by using quantum confinement that increases a number of surface mid-gap states[16] and quenches temperature-dependent structural fluctuations predominant in bulk crystals.[1]

A common feature of dynamic disorder is the onset of a wide Rayleigh wing[17] or a low-frequency Raman peak.[1,2] Nano-crystals (NCs) flare a spectrally broad high-energy emission that is red-shifted and size-dependent.[18] In spatially-confined metals, both emissions overlap completely and we observe a single band peaked at the pumping wavelength,[19] whereas these are spectrally well-resolved in disordered semiconductors.[3] The high-energy emission originates from inelastic light scattering by fluctuations of the electronic density in the vicinity of the Fermi level, as pioneered by A. Mal'shukov in 1989.[20] To date, this effect is understood as *electronic Raman scattering* (ERS) in which initial and final electronic states are different,



and optical transitions may be indirect due to the electron-photon momentum matching.[21–24] This is achieved by generating a near-field photon with increased momentum.[25] The ERS is similar to Compton effect[26] for visible radiation in which near-field photons are scattered by atomic-scale fluctuations of the electronic polarizability in polar crystals.

Metal halide perovskite crystals possess a specific $ABX_3$ structure in which the 4A and B cations interact with the halide 6X anions in such a way to form chemically stable 3D corner-sharing octahedral rigid frameworks $[BX_6]^{4-}$.[9] A direct bandgap semiconductor CsPbBr$_3$ ($E_g = 2.37$ eV) serves as a model system showing long-range chains of cross-linked $[PbBr_6]^{4-}$ octahedra. Though even considerable advances in understanding the mechanisms of emission in perovskite crystals,[27–30] there is still a number of fundamental concerns that continue to drive interest of the scientific community. These include (1) highly-emissive quantum dots in a glass matrix,[31–33] (2) PL blinking,[34–36] (3) enhanced PL when phase transitions,[27,37] (4) single-photon anti-Stokes photoluminescence (aS-PL) under sub-band pump,[38–40] (5) PL redshift in quantum dots,[18,28] (6) super-photoluminescence (super-PL) in cross-linked nanocrystals (or superlattices),[41–43] (7) single-photon superradiance in quantum dots,[44] (8) low-frequency Raman (disorder) peak[1,3] and (9) reversible crystal-glass transition in perovskites.[45]

In this article, we develop a physical model based on *electronic Raman scattering* (ERS) and address aforementioned challenges by using a concept of *glass-crystal-interface* which is a system of cross-linked CsPbBr$_3$ NCs and a host crystal.[3] In this model, the NCs capture the light and generate photo-electrons in the conduction band through the ERS, and then the charge-carriers tunnel into the crystal and recombine radiatively. This is claimed by PL/ERS blinking caused by thermal fluctuations of cross-linked $[PbBr_6]^{4-}$ octahedra. In addition, PL/ERS kinetics recognizes enhanced spontaneous PL and spontaneous bunching PL as soon as PL/ERS blinking is synchronized. The ERS process is an alternative mechanism for detrapping charge-carriers from mid-gap states into the conduction band when phonon energy is insufficient.



**DISCUSSION AND RESULTS**

Solution-processed synthesis of perovskite crystals is a traditional route to engineer static defects which are still difficult to control.[46,47] Our strategy aims to generate static/dynamic defects using a pulsed dc bias applied to a pristine $CsPbBr_3$ crystal pad, as-grown on a sapphire substrate (see details in Methods), as schematically shown in Figure 1 a. A PL study of $CsPbBr_3$ at 473 nm indicates the band-to-band transition at ca. 2.37 eV (523 nm) and confirms the spectral homogeneity across the crystal (Supplementary Information, Section I). A detailed procedure for integrating the $CsPbBr_3$ pad into the circuit in which electrodes are metallic-type single walled carbon nanotubes (SWCNTs) can be found in Refs.[48,49] Figure 1 b shows a scanning electron microscopy (SEM) image of a metal-semiconductor-metal (MSM) structure enabling switchable photovoltaic effect (Figure 1 c).[50,51] The photocurrent map exhibits negative (n-type) and positive (p-type) regions in the vicinity of the cathode and anode, respectively. Photogenerated charge-carries are separated by energy band bending due to Schottky contact, resulting the n-i-p junction in the MSM structure (Figure 1 d). The change in dc polarity when raster scanning a 633 nm focused laser beam (no bias) is due to the mirror symmetry of the band bending at the opposite (n- and p-type region) ends (Figure 1 c). Upon applying forward external dc bias, coupled ionic and electronic transport increases the concentration of bromine vacancies ($V_{Br}^+$) and bromine anions ($Br^-$) near the cathode and anode, respectively (Figure 1 c).[50] This creates a built-in electrical field that balances and weakens the external electrical field. The internal electrical field modifies a current kinetics when circuit-off noticeably, provided that the external bias was above 3 V (Figure S2 a).[49] Sharp downward current peaks appear as a response to the rapid relaxation of electrons invoked by the built-in electric field. This is precisely the voltage threshold at which a mixed ion-electron semiconductor can serve as either an electroluminescent photodetector or light-emitting diode (LED).[49] The dc current kinetics when circuit-on indicates a charging process of the n-i-p junction. Sporadic current oscillations at 9 V stem from light-enhanced electroluminescence



(EL) (Figure 1 e and Figure S2 b) which blink, as seen in movie 1. The EL occurs solely in the n-type region (Figure 1 c). In this region, the fast thermalization of hot electrons generates enough heat enabling to modify the sample chemically and structurally in those areas that shined (Figure S3 a-d). It is important to highlight that a green PL photon arising from interband radiative recombination is intrinsically spatially-confined and, therefore it has the expanded momentum, allowing indirect optical transitions to populate the conduction band excessively. Yet, one should mention current-induced Joule heating that competes with optical heating owing to parasitic re-absorption of the green emission. In general, these processes hold simultaneously and their separation is not a trivial task.[52,53]

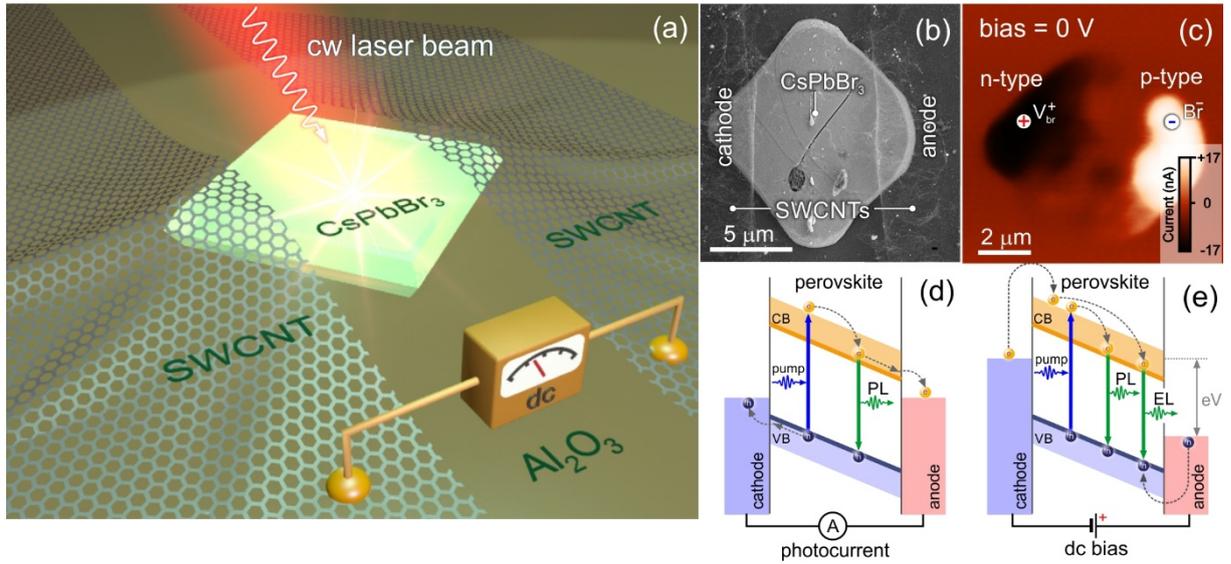

**Figure 1**. (a) Artistic illustration of a CsPbBr$_3$ pad, whose left and right corners are top-covered with SWCNT electrodes, (b) SEM image of the CsPbBr$_3$ perovskite pad (top view). (c) Photocurrent map across the perovskite pad under 633 nm cw illumination when dc bias off. (d) and (e) Sketches of photoluminescence mechanisms when external dc voltage bias off and on.

The multi-pulsed dc bias enables to create a disordered state or glass[45] that imbalances the primary distribution of charge-carriers. Destructuring stops when charging flips into



discharging, as resulting downward current kinetics (Figure S3 e). In this area, micro-sized CsPbBr$_3$ domains can transform into Cs$_4$PbBr$_6$ ($E_g = 3.9$ eV) NCs at temperatures above 350°C,[54] which are easily recognized by low-frequency Raman spectroscopy. Figure S4 shows Raman spectra of as-synthesized pristine CsPbBr$_3$ and Cs$_4$PbBr$_6$ for further analysis. Intense heating of shining areas of the perovskite crystal is supported by its low thermal conductivity (0.5 W m$^{-1}$K$^{-1}$),[55] making it difficult to scatter heat. As mentioned above, the temperature increase during EL emission is achieved by optical transitions from shallow/deep traps into the conduction band, followed by electron thermalization and heat release (Figure S3 c,d). The glass and crystal regions are separated by a 2D order-disorder transition region, further termed as a *glass-crystal-interface* (Figure 2 a), consisting of long-range coupled crystalline and nano-crystalline structures.

Figure 2 b shows dc-induced Raman spectra of CsPbBr$_3$ registered at the glass-crystal interface (red spot in Figure 2 a) exposed to a sub-band pump (632.8 nm, 1.959 eV) with the fluence of 0.4 MW/cm$^2$ for 10 s. Along with vibrational modes attributed to pristine CsPbBr$_3$ at 310 cm$^{-1}$ (2LO Pb-Br stretching mode)[56] and metallic SWCNTs at 1360 cm$^{-1}$ (D-mode), 1590 cm$^{-1}$ (G-mode) and 2720 cm$^{-1}$ (2D-mode),[57] we can observe three broad emissions: (a) single-photon anti-Stokes PL/EL, (b) low-energy electronic Raman scattering (*l*-ERS) and (c) high-energy electronic Raman scattering (*h*-ERS).[3] Both *l*-ERS and *h*-ERS, reconstructed with a regularized least-square method, originate from the electron-photon momentum matching owing to the increased momentum of a near-field photon generated by long-range structural fluctuations and quantum confinement, respectively. This is the result of the non-propagating optical field containing at least one imaginary wavenumber component.[58]

Overall, quantum confinement and long-range fluctuations can be characterized by a correlation length $l_c$ which is similar to excitonic delocalization extent.[44] In the first case, this is precisely the NC size, the latter determines a spatial extend for in-phase oscillations of rigid



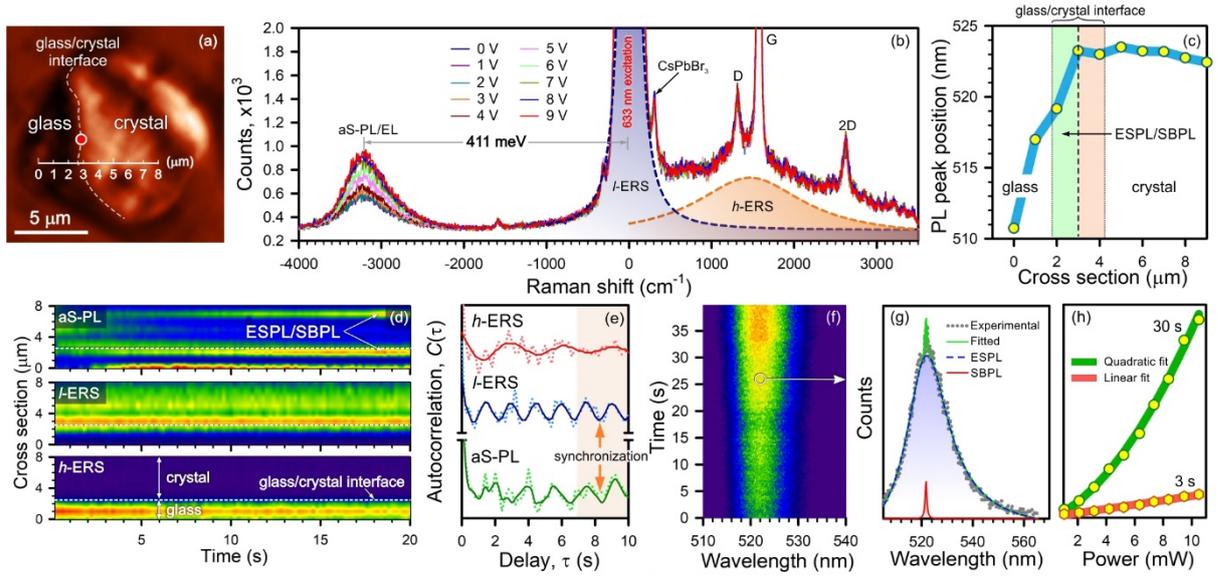

**Figure 2**. (a) A confocal optical image of a CsPbBr₃ pad consisting of glass and crystal regions separated with a glass-crystal interface. (b) Raman spectra of a CsPbBr₃ pad exposed to 633 nm excitation pump with the fluence of 0.4 MW/cm² and different dc bias, registered at the glass-crystal interface (red spot in Figure 2 a). (c) A plot of PL peak position vs cross section shown in Figure 2 a. (d) aS-PL, *l*-ERS and *h*-ERS kinetics along a cross section marked in Figure 2 a. (e) Autocorrelation function for three processes registered at the glass-crystal interface. (f) 1D Raman map vs time. (g) A PL spectrum at the spot marked in Figure 2 f and its numerical decomposition into the enhanced spontaneous PL (ESPL) (dashed blue curve) and spontaneous bunching PL (SBPL) (solid red curve) bands. (h) Pump-dependent PL intensity at the glass-crystal interface after 3 s and 30 s.

frameworks. When $l_c \ll \lambda$, this effect can be understood in terms of the dipole-dipole interaction between electronic and off-center ion polarizability.[2] This formalism explains the origin of a broad Rayleigh wing in highly associated liquids.[17,59] However, it is insufficient for a system possessing crystal-liquid duality[7] in which the correlation length can increase dramatically ($l_c \leq \lambda$). In this case, inelastic low-frequency scattering is associated with long-range structural oscillations rather than rotation and translation of coupled light molecules, as



previously thought.[59] This means that a central (disorder) peak observed in such systems is precisely the Raman scattering[1,2] or $l$-ERS[3] (Figure S5).

The central Raman peak is governed by the spontaneous long-range structural correlations that are sensitive to temperature. Upon cooling down to 80 K, this peak disappears completely for $CsPbBr_3$.[1] In addition, temperature-dependent phase transitions[27] can affect the dynamic disorder in $CsPbBr_3$ crystals. Figure S6 shows a temperature dependence of the Raman peak linewidth, termed as Urbach energy,[3] when heating/cooling within the range of $20 \div 160°C$. It is important to note that the Urbach energy exhibits an anomalous rise in the tetragonal β-phase when cooling, as resulting enhanced PL at the phase transition β → γ.[27,37] This phenomenon, that is still poorly understood, could be linked to the formation of longer structural chains through relaxing internal stresses at the $\alpha \to \beta$ transition (Figure S6).

Current-induced crystal-to-glass transformation occurs in regions which emit the EL, as shown in Figure S7 a. The $l$-ERS intensity is independent of dc bias when EL off and increases otherwise. In glass, the EL intensity decreases due to the chemical transformation of $CsPbBr_3$ to $Cs_4PbBr_6$ and the formation of NCs, suppressing the $l$-ERS intensity (Figure S7 b). This means that quantum confinement prevents long-range structural correlations that allow inelastic scattering of incident light. On the other hand, NCs generate larger momentum in magnitude near-field photons that intensify optical transitions from mid-gap states to the conduction band, resulting $h$-ERS (Figure S8). The heavy $h$-ERS tail extending beyond 3000 cm$^{-1}$ (>0.4 eV)[3] indicates the formation of tiny NCs with a huge density of surface mid-gap states. In contrast to $l$-ERS, the $h$-ERS band is red-shifted and its intensity increases noticeably when moving from crystal to glass (Figure S8).

The electron-photon momentum matching governs both the processes, $l$-ERS and $h$-ERS, through long-range structural fluctuations in crystals and quantum confinement in NCs, respectively. The vibrational Raman intensity defined as $I_{nm} \sim \langle \vec{\alpha}_{nn}^* \vec{\alpha}_{mm} \rangle$ (where $\vec{\alpha}$ is an atomic polarizability tensor for two quantum states $\psi_n$ and $\psi_m$, $\langle \cdots \rangle$ is an averaging symbol) can be



modified for the ERS intensity, which is a function of correlation length $l_c$ or wavenumber $k_0$, as follows

$$I_{nm}(\omega, k_0) = \sigma_{\pm}(\omega \pm \omega_{nm})^4 \int_{k_0-\delta k}^{k_0+\delta k} \langle \vec{\alpha}_{nn}^*(\mathbf{k}) \vec{\alpha}_{mm}(\mathbf{k}) \rangle e^{-\left(\frac{k}{2\delta k}\right)^2} d\mathbf{k}, \qquad (1)$$

where $\sigma_{\pm}$ are the cross-sections for anti-Stokes and Stokes scattering, the atomic polarizability taking into account the phase of incident radiation when $r \sim \lambda$ has the following view

$$\vec{\alpha}_{nn}(\mathbf{k}) = \frac{1}{h} \sum_p \frac{\mathbf{D}_{np}(\mathbf{k}) \mathbf{D}_{pn}(\mathbf{k})}{\omega_{pn} - \omega + i\Gamma} + \frac{\mathbf{D}_{pn}(\mathbf{k}) \mathbf{D}_{np}(\mathbf{k})}{\omega_{pn} + \omega - i\Gamma}, \qquad (2)$$

where $\mathbf{D}_{nm}(\mathbf{k}) = \langle \psi_n | e^{i\mathbf{k}\mathbf{r}} \, \partial/\partial \mathbf{r} \, | \psi_m \rangle$ is the transient electrical dipole moment between electronic states $\psi_n$ and $\psi_m$, $\Gamma$ is the width of electronic level. The momentum of a confined photon fluctuates around $k_0 \cong \pi/l_c$ within $\delta k$.[22,23] The uncertainty $\delta k$ follows from the Heisenberg principle due to quantum confinement. By this reason, $\langle \vec{\alpha}_{nn}^*(\mathbf{k}) \vec{\alpha}_{mm}(\mathbf{k}) \rangle$ is averaged using the Gaussian momentum distribution in $k$-space. In the long-wavelength approximation $r \ll \lambda$, a quantum system can be seen as a set of quantum oscillators with electrical dipole moments $\mathbf{D}_{nm} e^{i\omega_{nm}t}$, in other words, the phase of light wave is constant for all oscillators. Otherwise, it is necessary to take into account the phase allowing to perceive the quantum system as currents and charges distributed in space.

Unlike the *l*-ERS and *h*-ERS bands which are not sensitive to applied external voltage (Figure 2 b), the aS-PL/EL intensity increases at the glass-crystal-interface exposed to both dc bias and sub-band pump. A single-photon up-conversion follows from the linear behavior of the pump-dependent PL intensity, as shown in Figure S9. Despite partial destruction, the crystal retains its luminescent properties, but the PL blueshift indicates the presence of NCs in the glass (Figure 2 c).[60] This plot clearly visualizes the glass-crystal-interface of 2 μm in width (axial size) along the scale bar depicted in Figure 2 a. aS-PL, *l*-ERS and *h*-ERS kinetics along the scale bar shed light on the mechanisms of emission from cross-linked crystalline and nano-crystalline structures at the glass-crystal-interface (Figure 2 d). The kinetics maps were build up with the temporal and spatial resolution of 200 ms and 500 nm, respectively. Through the



entire crystal, the aS-PL intensity drops in time, except the glass-crystal-interface and a region at 7 μm, in which it starts to grow up in 6-8 s. The fact that the *l*-ERS intensity is time-independent in crystal, disappears in glass and increases at the glass-crystal-interface confirms the above hypothesis about long-range structural fluctuations. Yet, this concept is supported by the absence of *h*-ERS in crystal and the downward kinetics in glass. A calculation of autocorrelation function, $C(\tau) = \int I(t)I(t - \tau)dt$, claims that all three emissions blink with the characteristic period: aS-PL – 1.9 s, *l*-ERS – 1.5 s and *h*-ERS – 3 s (Figure 2 e). A close inspection of the autocorrelation for aS-PL indicates two periods: 1.5 and 1.9 s. Importantly, there is no a specific PL blinking period typically linked to the detection threshold.[35] The increase in the aS-PL intensity over time is associated with its reabsorption, saturating the crystal conduction band with free charge-carriers (Figure S10). The periodic oscillations of the *l*-ERS intensity in the crystal indicate thermal fluctuations of a cross-linked $[PbBr_6]^{4-}$ octahedra network (Figure 2 e). The decay kinetics of the *h*-ERS intensity results from the depleted population of charge-carriers residing the surface mid-gap states in NCs. Once this process terminates both the aS-PL and the *l*-ERS begin to synchronize with the period of 1.5 s, and we observe enhanced spontaneous PL (ESPL) and sporadic narrow peaks (spontaneous bunching PL (SBPL)) simultaneously, as follows from 1D aS-PL kinetics (Figure 2 f and Figure S10). The SBPL process closely resembles amplified spontaneous emission,[61] an important phenomenon for mirrorless lasing at room temperature.[62]

The ESPL is driven by the free-carrier recombination by which all emitted photons are incoherent. The narrow peaks, randomly distributed over the ESPL band (Figure 2 f), appear due to the avalanche optical transitions of bunched charge-carriers emitting coherent photons. However, the photons emitted by different bunches are incoherent. This bunching effect can be explained by a large polaron formation[7] in the crystal surrounded by a huge number of NCs being a charge-carrier reservoir (Figure S11). This glass-crystal interface might likely enable true random lasing[63] when cooling below 100 K.[62]



The aS-PL band, observed in our experiment, is decomposed into the elementary peaks, the broad ASE band (the product of the Lorentzian profile and the exponential function taking into account the Urbach tail) and the Lorentzian-shaped narrow BSE band, using a regularized least-square method (Figure 2 g). The pump-dependent aS-PL intensity measured in 3 and 30 s exhibits linear and quadratic behaviors. The latter excludes two-photon absorption (Figure S9), and indicates the nonlinear regime at the glass-crystal interface with the temporal threshold of 6-8 s (Figure 2 h). Initially, the excitonic radiative recombination is predominant owing to the lower concentration of free charge-carriers in the crystal exposed to the sub-band pump. Once the re-absorbed green PL saturates the conduction band of the crystal with free charge-carriers, the radiative recombination exhibits the electron-hole (quadratic) behavior (Figure 2 h and Figure S9).[64,65] These effects are basically played in glass within the transition zone, as marked in Figure 2 c. Time-delayed synchronization of the aS-PL and $l$-ERS blinks (Figure 2 e) is achieved by using the sporadic green PL as a feedback. Unlike conventional perovskite crystals, the glass-crystal interface contains an inexhaustible reservoir of trapped charge-carriers in NCs, which are detrapped by using the $l$-ERS mechanism and moved into the crystal conduction band (Figure S10).

A total blinking period $\tau$ of aS-PL at the glass/crystal interface is determined by the sum of the charge-carrier lifetime $\tau_{NC}$ within the NC bandgap, the nanocrystal-to-crystal transition (tunneling) time $\tau_{NC \to C}$ and the lifetime of an electronic level in the conduction band edge $\tau_C$: $\tau = \tau_{NC} + \tau_{NC \to C} + \tau_C$. Since $\tau_{NC} \gg \tau_{NC \to C} > \tau_C$, it is sufficient to estimate $\tau_{NC}$ per unit volume when an electron resides the Urbach bridge:

$$\tau_{NC} \cong E_g \frac{dn}{dE} N \tau_0, \tag{3}$$

where $N$ is the concentration of static defects (estimated to be $10^{15} \div 10^{16}$ cm$^{-3}$)[35], $E_g$ is the bandgap energy (2.37 eV for CsPbBr$_3$), $\tau_0$ is a period of thermal fluctuations of [PbBr$_6$]$^{4-}$ octahedra ($\tau_0 \approx 1 \div 10$ ps)[40,66], and $dn/dE$ is the density of dynamic states across the Urbach bridge, defined as



$$\frac{dn}{dE} \sim \frac{1}{e}\left(\frac{l_c}{\lambda}\right)^d, \qquad (4)$$

here $e$ is the charge of an electron, $\lambda$ is the pumping wavelength (632.8 nm), $d$ is the topological

dimension ($d = 3$ for bulk crystal). In the range of $l_c = 1 \div 100$ nm (the upper limit is driven

by the lateral size of twin domains, measured with TERS microscopy (Figure S12), the blinking

period varies between microseconds and tens of seconds, which is consistent with earlier

theoretical predictions.[35] This characteristic time can monitor with the naked eye within a few

seconds, as seen in movie 1 - the EL sporadically emits when applied dc bias of 9 V.

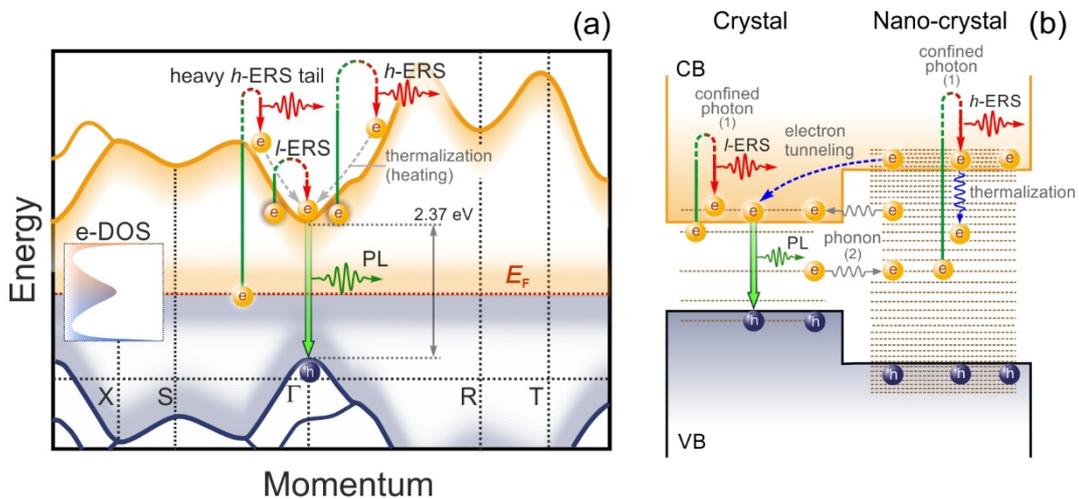

**Figure 3**. (a) Energy band structure of CsPbBr$_3$. (b) Schematic illustration of optical transitions
at the glass-crystal interface (not to scale).

Using the ERS concept, we suggest an alternative mechanism of up-conversion

photoluminescence in metal halide perovskites under sub-band pump.[38,47,67–69] One of the

fundamental barriers to explain anti-Stokes shifts of a few hundred meV is insufficient energy

of phonons to move charge-carriers from the deep states to the conduction band. Bo Wu et al

interpreted this effect using thermal fluctuations of coupled rigid frameworks making the

bandgap to oscillate.[40] As a result, the charge-carriers can tunnel from traps to the conduction

band. In our model, shown in Figure 3 a, the anti-Stokes shift by 411 meV (Figure 2 b) can be

interpreted through the *l*-ERS in crystal and the *h*-ERS in NC, in which the momentum of an



incident photon is expanded either by long-range structural oscillations of $[PbBr_6]^{4-}$ octahedra and quantum confinement, respectively. It is important to stress that the *l*-ERS process enables to drag trapped electrons from shallow states into the conduction band predominantly, whereas the *h*-ERS process moves the electrons from all traps within the forbidden gap. Those charge-carriers that transfer from the mid-gap states into the conduction band contribute to the heavy *h*-ERS tail (Figure 3 a and Figure S8).

Let us consider a model system of coupled crystal and nano-crystal structures to explain the enhancement of the emissions at the glass-crystal interface upon sub-band pump (Figure 3 b). There are two mechanisms making the trapped electrons to travel between the localized and extended states: (1) electron-photon interaction[3] and (2) electron-phonon interaction.[70,71] While phonon-assisted indirect optical transitions are insufficient to observe inelastic broadband emissions, a concept of confined photon with increased momentum can shed light on the origin of observable emissions. Sub-wavelength light localization in crystal, driven by long-range structural fluctuations, increases photon momentum insignificantly. This leads to temperature-dependent *l*-ERS throughout the crystal, but this process disappears in glass due to quantum confinement (Figure 2 d). This is precisely the emission mechanism that is responsible for PL blinking. It is important to stress that the density of dynamic states in a perovskite crystal exceeds that of static states. The picture is opposite in NCs wherein the density of surface static states is maximal. The highly confined photon initiates indirect optical transitions from the mid-gap states to the conduction band, which predominantly occur in glass and emit the *h*-ERS (Figure 2 d). This emission is red-shifted by a few hundred meV, and previously interpreted as PL.[16] An electron in the NC conduction band is either thermalized back to populate a static defect state or tunnel into the crystal conduction band for radiative recombination. The green PL is subsequently reused for band-to-band pumping.[72] This mechanism allows a PL broadening and blueshift near the glass-crystal interface (Figure S13). The enhanced *l*-ERS process (Figure 2 d) can greatly pump the conduction band using confined photons, further



followed by excess radiative recombination in the band-to-band region (Figure S13 e). In this figure, we see the inhomogeneous broadening of the PL band caused by the larger cross-section of band-to-band absorption in contrast to that of the sub-band absorption near the glass-crystal interface. Photon-momentum-enabled interband optical transitions drive the PL blueshift that previously perceived as photoluminescence of quantum dots formed in the crystal (Figure S13 e).[32,33]

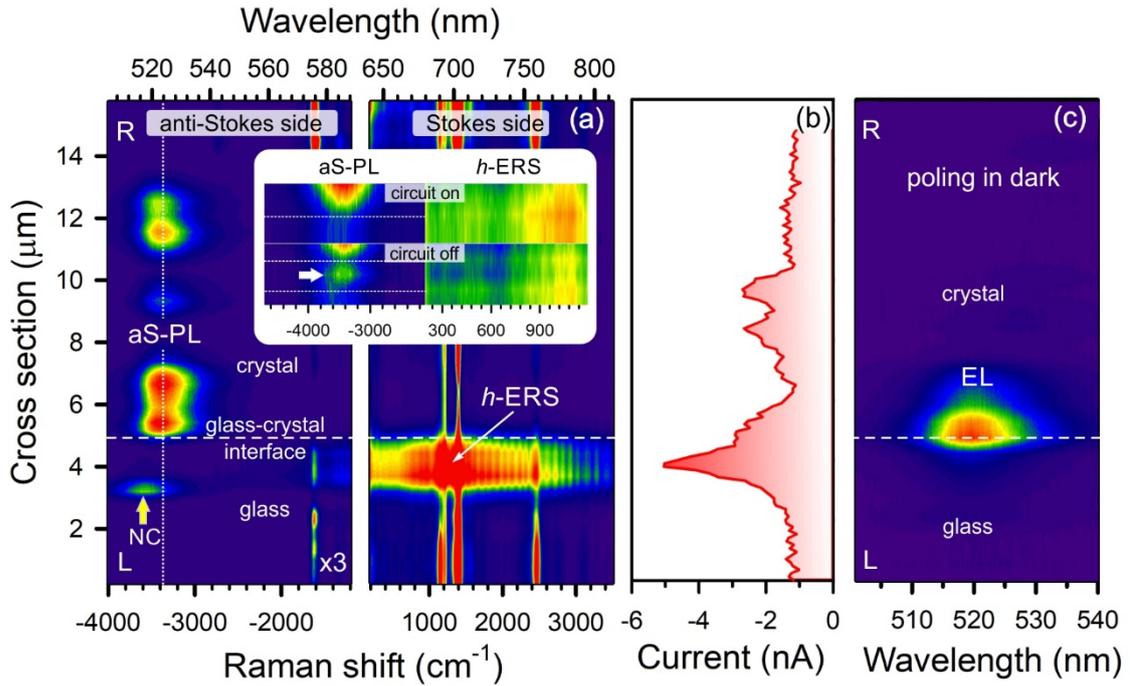

**Figure 4**. (a) 1D Raman intensity map, (b) a photocurrent distribution curve and (c) 1D EL map at dc bias of 7 V along the cross section highlighted in Figure 2 a. The inset in the left panel shows the aS-PL and *h*-ERS at the glass-crystal-interface when the circuit is on and off. The dotted straight lines mark the *h*-ERS maximum. The arrow indicates an additional aS-PL flare while circuit off.

Figure 4 a shows a 1D Raman intensity map across the CsPbBr₃ crystal pad exposed to sub-band pump at 633 nm. This map clearly confirms the physical model in Figure 3 b. The *h*-ERS and aS-PL are enhanced near the glass-crystal-interface in glass and crystal, respectively. The intensity of the anti-Stokes side was increased by a factor of 3 for easy comparison of the



anti-Stokes and Stokes signals in Figure 4 a. In glass, we find a NC emitting the PL at 516 nm, as marked with the arrow. The enhancement of the *h*-ERS is accompanied by an increase in current (Figure 4 b), which indicates an excess of photo-generated electrons in the conduction band. Upon moving from glass to crystal, the current drops due to the tunneling of photo-generated electrons into the crystal wherein they recombine to emit the aS-PL. This observation holds the above hypothesis that the *h*-ERS process moves charge-carriers from traps within the bandgap into the conduction band (Figure 3 b). While the circuit is on, the *h*-ERS signal is maximum and the aS-PL is minimum at the glass-crystal-interface, as seen from the inset in Figure 4 a. Once the circuit is turned off, excess photo-generated free charge-carriers begin to tunnel into the crystal, and we observe an extra aS-PL flare, marked with the arrow in the inset (Figure 4 a). This is a strong argument in favor of enrichment of the conduction band with free charge-carries transferred from the Urbach bridge using the *h*-ERS process. In contrast, the *l*-ERS process, not shown in Figure 4 a, is insensitive to circuit switching (Figure S14), since most charge-carriers remain to be trapped within the bandgap when soft inelastic light scattering (Figure 3 b).

Figure 4 c shows the EL generation near the glass-crystal interface in dark when external applied voltage of 7 V. Unlike the initial crystal that sporadically emits the EL in the n-type region, the formation of a glass-crystal-interface allows the n-i-p junction to be tuned for radiative recombination.

Thus, this result can be used for producing 3D ordered bulk systems to improve emission efficiency, for example, as seen from recently synthesized multicomponent nanocrystal superlattices.[73] Reversible glass-to-crystal transitions[45] in systems with crystal-liquid duality[7] may serve as a promising platform for neuromorphic computing.[74] The concept of expanded photon momentum allows not only explain the origin of observable inelastic broadband emissions, but also develop next-generation optoelectronic technologies and devices.



**CONCLUSION**

In this work, we have developed a physical model based on *electronic Raman scattering* to address the enhanced emissions in a metal halide perovskite that is a system with crystal-liquid duality. This model includes a new concept of *the glass-crystal-interface* that enables to unravel a mechanism of an anomalous anti-Stokes shift by 411 meV for $CsPbBr_3$. The incident photon is scattered by long-range structural fluctuations of cross-linked $[PbBr_6]^{4-}$ octahedra or spatially-confined nano-crystals, which increase its momentum. This leads to elastic/soft inelastic low-frequency ERS in crystal and hard inelastic high-energy ERS in NCs. The electron-photon momentum matching allows optical transitions from shallow/deep states of the Urbach bridge into the conduction band, further followed by charge-carrier tunneling from NCs to crystal for emitting up-conversion photoluminescence under sub-band pumping. We experimentally revealed PL/ERS blinking in a $CsPbBr_3$ crystal, directly associated with thermal oscillations of cross-linked $[PbBr_6]^{4-}$ frameworks which vanish upon cooling.[35] The *h*-ERS blinking decays rapidly enough (a few seconds) due to the depletion of electrons in the mid-gap states, and then we observe the synchronization of *l*-ERS and aS-PL, as resulting both enhanced spontaneous PL and spontaneous bunching PL. Electron transitions from mid-gap states to the conduction band is confirmed by the formation of the heavy *h*-ERS tail (above 0.4 eV) upon sub-band pump. The results, obtained in this study, will be beneficial in diverse fields of optoelectronic and photovoltaic applications, including (a) intrinsic white light-emitting diodes, (b) solar cells based on cross-linked amorphous-crystalline semiconductors, (c) Urbach-energy-based temperature sensors, and (d) structural analysis of defects in disordered solids. Lastly, our findings highlight the role of expanded photon momentum in inelastic broadband emission from nanostructured solids and media with crystal-liquid duality.



**ACKNOWLEDGEMENT**

Authors thank Prof. Alexander Fishman and Dr. Alexey Noskov for the fruitful discussions. This work was granted by from the subsidy allocated to Kazan Federal University for the state assignment in the sphere of scientific activities (FZSM-2022-0021) and the Kazan Federal University Strategic Academic Leadership Program (PRIORITY-2030). All authors acknowledge a technical support from NT-MDT BV (The Netherlands).

**METHODS**

*Synthesis of a CsPbBr$_3$ pad*

CsPbBr$_3$ microplates on a sapphire substrate were synthesized by high-temperature sublimation similar to the previously reported.[49] Firstly, the source glass substrate was prepared using solution-processed synthesis. Then, the target sapphire substrate was cleaned, consistently, in water, isopropanol and acetone for 5 min. After that, the two substrates were placed in the high-temperature titanium hotplate (PZ 28-3TD) with the 6 mm gap between them. The initial temperature was 250ºC. While the synthesis, the temperature, first, raised to 530 ºC during 15 min, then, it held for 10 min, and, lastly, followed by lowering down to 485ºC during 10 min. As a result, CsPbBr$_3$ microplates were sublimated on the sapphire substrate.

*Metal-semiconductor-metal structure fabrication*

A single-walled carbon nanotube (SWCNT) thin film (20 nm thickness) was dry-transferred on top of the as-grown CsPbBr$_3$ microplates. To finish the device fabrication, a light conversion Pharos femtosecond laser (200 fs pulse duration, wavelength 1030 nm) was used to cut the SWCNT film into two pieces above the chosen perovskite microplate. In order to provide micrometer sized distance between the two SWCNT electrodes, the light was focused through a 50×lens (NIR Mitutoyo, NA=0.65), while the xy-position was altered with a programmable piezo controller (Standa). For the successful ablation, the laser fluence level was



controlled to be below 0.1 J/cm$^2$. For more details on the process of SWCNT ablation see references [48-49]

*Atomic force microscopy*

The multimode scanning probe microscope NTEGRA PRIMA (NT-MDT) was utilized for visualizing a topography of the CsPbBr$_3$ pad surface. The AFM probes of the "VIT_P" series with resonant frequencies around 350 kHz were used in AFM measurements. The CsPbBr$_3$ pad was measured in tapping mode with a free amplitude $A_0$ of 10-20 nm and a set-point value of $A_0/2$.

*Far- and near-field Raman spectroscopy and microscopy*

Raman spectra and maps were captured with a multi-purpose analytical instrument NTEGRA SPECTRA™ (NT-MDT) in upright configuration. The confocal spectrometer was wavelength calibrated with a crystalline silicon (100) wafer by registering the first-order Raman band at 521 cm$^{-1}$. A sensitivity of the spectrometer was as high as ca. 1700 photon counts per 0.1 s provided that we used a 100× objective (N.A.=0.7), an exit slit (pinhole) of 100 μm and a linearly polarized light with the wavelength of 632.8 nm and the power at the sample of 10 mW. No signal amplification regimes of a Newton EMCCD camera (ANDOR) was used. Low-frequency Raman measurements were performed using a 633 nm Bragg notch filter (OptiGrate) with a spectral blocking window of 10 cm$^{-1}$.




**REFERENCES**

(1)     Yaffe, O.; Guo, Y.; Tan, L. Z.; Egger, D. A.; Hull, T.; Stoumpos, C. C.; Zheng, F.;

        Heinz, T. F.; Kronik, L.; Kanatzidis, M. G.; Owen, J. S.; Rappe, A. M.; Pimenta, M. A.;

        Brus, L. E. Local Polar Fluctuations in Lead Halide Perovskite Crystals. *Phys. Rev. Lett.*

        **2017**, *118*, 136001.

(2)     Vugmeister, B. E.; Yacoby, Y.; Toulouse, J.; Rabitz, H. Second-Order Central Peak in

        the Raman Spectra of Disordered Ferroelectrics. *Phys. Rev. B* **1999**, *59*, 8602–8606.

(3)     Kharintsev, S. S.; Battalova, E. I.; Noskov, A. I.; Merham, J.; Potma, E. O.; Fishman, D.

        A. Photon-Momentum-Enabled Electronic Raman Scattering in Silicon Glass. *ACS*

        *Nano* **2024**, *18*, 9557–9565.

(4)     Cohen, A. V.; Egger, D. A.; Rappe, A. M.; Kronik, L. Breakdown of the Static Picture

        of Defect Energetics in Halide Perovskites: The Case of the Br Vacancy in $CsPbBr_3$. *J.*

        *Phys. Chem. Lett.* **2019**, *10*, 4490–4498.

(5)     Wang, B.; Chu, W.; Wu, Y.; Casanova, D.; Saidi, W. A.; Prezhdo, O. V. Electron-Volt

        Fluctuation of Defect Levels in Metal Halide Perovskites on a 100 ps Time Scale. *J.*

        *Phys. Chem. Lett.* **2022**, *13*, 5946–5952.

(6)     Wu, B.; Yuan, H.; Xu, Q.; Steele, J. A.; Giovanni, D.; Puech, P.; Fu, J.; Ng, Y. F.;

        Jamaludin, N. F.; Solanki, A.; Mhaisalkar, S.; Mathews, N.; Roeffaers, M. B. J.;

        Grätzel, M.; Hofkens, J.; Sum, T. C. Indirect Tail States Formation by Thermal-Induced

        Polar Fluctuations in Halide Perovskites. *Nat. Commun.* **2019**, *10*, 484.

(7)     Miyata, K.; Atallah, T. L.; Zhu, X. Lead Halide Perovskites: Crystal-Liquid Duality,

        Phonon Glass Electron Crystals, and Large Polaron Formation. *Sci. Adv.* **2017**, *3*,

        e1701469.

(8)     Dey, A.; Ye, J.; De, A.; Debroye, E.; Ha, S. K.; Bladt, E.; Kshirsagar, A. S.; Wang, Z.;

        Yin, J.; Wang, Y.; Quan, L. N.; Yan, F.; Gao, M.; Li, X.; Shamsi, J.; Debnath, T.; Cao,

        M.; Scheel, M. A.; Kumar, S.; Steele, J. A.; et al. State of the Art and Prospects for





Halide Perovskite Nanocrystals. *ACS Nano* **2021**, *15*, 10775–10981.

(9)     Alaei, A.; Circelli, A.; Yuan, Y.; Yang, Y.; Lee, S. S. Polymorphism in Metal Halide
        Perovskites. *Mater. Adv.* **2021**, *2*, 47–63.

(10)    Zhang, D.; Shah, D.; Boltasseva, A.; Gogotsi, Y. MXenes for Photonics. *ACS Photonics*
        **2022**, *9*, 1108–1116.

(11)    Zhou, J.; Dahlqvist, M.; Björk, J.; Rosen, J. Atomic Scale Design of MXenes and Their
        Parent Materials – From Theoretical and Experimental Perspectives. *Chem. Rev.* **2023**,
        *123*, 13291–13322.

(12)    Novoselov, K. S.; Mishchenko, A.; Carvalho, A.; Castro Neto, A. H. 2D Materials and
        van der Waals Heterostructures. *Science* **2016**, *353*, aac9439.

(13)    Ma, L.-L.; Li, C.-Y.; Pan, J.-T.; Ji, Y.-E.; Jiang, C.; Zheng, R.; Wang, Z.-Y.; Wang, Y.;
        Li, B.-X.; Lu, Y.-Q. Self-Assembled Liquid Crystal Architectures for Soft Matter
        Photonics. *Light: Sci. Appl.* **2022**, *11*, 270.

(14)    Oses, C.; Toher, C.; Curtarolo, S. High-Entropy Ceramics. *Nat. Rev. Mater.* **2020**, *5*,
        295–309.

(15)    Strandell, D.; Wu, Y.; Mora-Perez, C.; Prezhdo, O.; Kambhampati, P. Breaking the
        Condon Approximation for Light Emission from Metal Halide Perovskite Nanocrystals.
        *J. Phys. Chem. Lett.* **2023**, *14*, 11281–11285.

(16)    Mooney, J.; Krause, M. M.; Saari, J. I.; Kambhampati, P. Challenge to the Deep-Trap
        Model of the Surface in Semiconductor Nanocrystals. *Phys. Rev. B* **2013**, *87*, 081201.

(17)    Gochiyaev, V. Z.; Malinovsky, V. K.; Novikov, V. N.; Sokolov, A. P. Structure of the
        Rayleigh Line Wing in Highly Viscous Liquids. *Philos. Mag. B* **1991**, *63*, 777–787.

(18)    de Boer, W. D. A. M.; Timmerman, D.; Dohnalová, K.; Yassievich, I. N.; Zhang, H.;
        Buma, W. J.; Gregorkiewicz, T. Red Spectral Shift and Enhanced Quantum Efficiency
        in Phonon-Free Photoluminescence from Silicon Nanocrystals. *Nat. Nanotechnol.* **2010**,
        *5*, 878–884.



(19)    Baffou, G. Anti-Stokes Thermometry in Nanoplasmonics. *ACS Nano* **2021**, *15*, 5785–5792.

(20)    Gass, A. N.; Kapusta, O. I.; Klimin, S. A.; Mal'shukov, A. G. The Nature of the Inelastic Background in Surface Enhanced Raman Scattering Spectra of Coldly-Deposited Silver Films. The Role of Active Sites. *Solid State Commun.* **1989**, *71*, 749–753.

(21)    Barnett, S. M.; Harris, N.; Baumberg, J. J. Molecules in the Mirror: How SERS Backgrounds Arise From the Quantum Method of Images. *Phys. Chem. Chem. Phys.* **2014**, *16*, 6544–6549.

(22)    Hugall, J. T.; Baumberg, J. J. Demonstrating Photoluminescence from Au is Electronic Inelastic Light Scattering of a Plasmonic Metal: The Origin of SERS Backgrounds. *Nano Lett.* **2015**, *15*, 2600–2604.

(23)    Baumberg, J. J.; Esteban, R.; Hu, S.; Muniain, U.; Silkin, I. V.; Aizpurua, J.; Silkin, V. M. Quantum Plasmonics in Sub-Atom-Thick Optical Slots. *Nano Lett.* **2023**, *23*, 10696–10702.

(24)    Inagaki, M.; Isogai, T.; Motobayashi, K.; Lin, K.-Q.; Ren, B.; Ikeda, K. Electronic and Vibrational Surface-Enhanced Raman Scattering: from Atomically Defined Au(111) and (100) to Roughened Au. *Chem. Sci.* **2020**, *11*, 9807–9817.

(25)    Bharadwaj, P.; Deutsch, B.; Novotny, L. Optical Antennas. *Adv. Opt. Photonics* **2009**, *1*, 438–483.

(26)    Compton, A. H. A Quantum Theory of the Scattering of X-rays by Light Elements. *Phys. Rev.* **1923**, *21*, 483−502.

(27)    Kharintsev, S. S.; Battalova, E. I.; Mukhametzyanov, T. A.; Pushkarev, A. P.; Scheblykin, I. G.; Makarov, S. V.; Potma, E. O.; Fishman, D. A. Light-Controlled Multiphase Structuring of Perovskite Crystal Enabled by Thermoplasmonic Metasurface. *ACS Nano* **2023**, *17*, 9235–9244.





(28)    Rainò, G.; Yazdani, N.; Boehme, S. C.; Kober-Czerny, M.; Zhu, C.; Krieg, F.; Rossell, M. D.; Erni, R.; Wood, V.; Infante, I.; Kovalenko, M. V. Ultra-Narrow Room-Temperature Emission from Single CsPbBr$_3$ Perovskite Quantum Dots. *Nat. Commun.* **2022**, *13*, 2587.

(29)    Hoffman, A. E. J.; Saha, R. A.; Borgmans, S.; Puech, P.; Braeckevelt, T.; Roeffaers, M. B. J.; Steele, J. A.; Hofkens, J.; Van Speybroeck, V. Understanding the Phase Transition Mechanism in the Lead Halide Perovskite CsPbBr$_3$ via Theoretical and Experimental GIWAXS and Raman Spectroscopy. *APL Mater.* **2023**, *11*, 041124.

(30)    Tailor, N. K.; Kar, S.; Mishra, P.; These, A.; Kupfer, C.; Hu, H.; Awais, M.; Saidaminov, M.; Dar, M. I.; Brabec, C.; Satapathi, S. Advances in Lead-Free Perovskite Single Crystals: Fundamentals and Applications. *ACS Materials Lett.* **2021**, *3*, 1025–1080.

(31)    Biswas, K. Revisiting the Origin of Green Emission in Cs$_4$PbBr$_6$. *Mater. Adv.* **2022**, *3*, 6791–6798.

(32)    Wang, C.; Wang, Y.; Su, X.; Hadjiev, V. G.; Dai, S.; Qin, Z.; Calderon Benavides, H. A.; Ni, Y.; Li, Q.; Jian, J.; Kamrul Alam, Md.; Wang, H.; Robles Hernandez, F. C.; Yao, Y.; Chen, S.; Yu, Q.; Feng, G.; Wang, Z.; Bao, J. Extrinsic Green Photoluminescence from the Edges of 2D Cesium Lead Halides. *Adv. Mater.* **2019**, *31*, 1902492.

(33)    Quan, L. N.; Quintero-Bermudez, R.; Voznyy, O.; Walters, G.; Jain, A.; Fan, J. Z.; Zheng, X.; Yang, Z.; Sargent, E. H. Highly Emissive Green Perovskite Nanocrystals in a Solid State Crystalline Matrix. *Adv. Mater.* **2017**, *29*, 1605945.

(34)    Seth, S.; Podshivaylov, E. A.; Li, J.; Gerhard, M.; Kiligaridis, A.; Frantsuzov, P. A.; Scheblykin, I. G. Presence of Maximal Characteristic Time in Photoluminescence Blinking of MAPbI$_3$ Perovskite. *Adv. Energy Mater.* **2021**, *11*, 2102449.

(35)    Gerhard, M.; Louis, B.; Camacho, R.; Merdasa, A.; Li, J.; Kiligaridis, A.; Dobrovolsky,



A.; Hofkens, J.; Scheblykin, I. G. Microscopic Insight into Non-Radiative Decay in Perovskite Semiconductors from Temperature-Dependent Luminescence Blinking. *Nat. Commun.* **2019**, *10*, 1698.

(36)     Yuan, H.; Debroye, E.; Caliandro, G.; Janssen, K. P. F.; van Loon, J.; Kirschhock, C. E. A.; Martens, J. A.; Hofkens, J.; Roeffaers, M. B. J. Photoluminescence Blinking of Single-Crystal Methylammonium Lead Iodide Perovskite Nanorods Induced by Surface Traps. *ACS Omega* **2016**, *1*, 148–159.

(37)     Dobrovolsky, A.; Merdasa, A.; Unger, E. L.; Yartsev, A.; Scheblykin, I. G. Defect-Induced Local Variation of Crystal Phase Transition Temperature in Metal-Halide Perovskites. *Nat. Commun.* **2017**, *8*, 34.

(38)     Ma, X.; Pan, F.; Li, H.; Shen, P.; Ma, C.; Zhang, L.; Niu, H.; Zhu, Y.; Xu, S.; Ye, H. Mechanism of Single-Photon Upconversion Photoluminescence in All-Inorganic Perovskite Nanocrystals: The Role of Self-Trapped Excitons. *J. Phys. Chem. Lett.* **2019**, *10*, 5989–5996.

(39)     Zhu, D.; Sun, Y.; Yuan, S.; Gao, R.; Wang, Y.; Ai, X.-C.; Zhang, J.-P. Intragap State Engineering for Tunable Single-Photon Upconversion Photoluminescence of Lead Halide Perovskite. *J. Phys. Chem. C* **2022**, *126*, 2447–2453.

(40)     Wu, B.; Wang, A.; Fu, J.; Zhang, Y.; Yang, C.; Gong, Y.; Jiang, C.; Long, M.; Zhou, G.; Yue, S.; Ma, W.; Liu, X. Uncovering the Mechanisms of Efficient Upconversion in Two-Dimensional Perovskites with Anti-Stokes Shift up to 220 meV. *Sci. Adv.* **2023**, *9*, eadi9347.

(41)     Biliroglu, M.; Findik, G.; Mendes, J.; Seyitliyev, D.; Lei, L.; Dong, Q.; Mehta, Y.; Temnov, V. V.; So, F.; Gundogdu, K. Room-Temperature Superfluorescence in Hybrid Perovskites and Its Origins. *Nat. Photonics* **2022**, *16*, 324–329.

(42)     Rainò, G.; Becker, M. A.; Bodnarchuk, M. I.; Mahrt, R. F.; Kovalenko, M. V.; Stöferle, T. Superfluorescence from Lead Halide Perovskite Quantum Dot Superlattices. *Nature*





**2018**, *563*, 671–675.

(43)     Russ, B.; Eisler, C. N. The Future of Quantum Technologies: Superfluorescence from
         Solution-Processed, Tunable Materials. *Nanophotonics* **2024**, in press. DOI:
         10.1515/nanoph-2023-0919

(44)     Zhu, C.; Boehme, S. C.; Feld, L. G.; Moskalenko, A.; Dirin, D. N.; Mahrt, R. F.;
         Stöferle, T.; Bodnarchuk, M. I.; Efros, A. L.; Sercel, P. C.; Kovalenko, M. V.; Rainò, G.
         Single-Photon Superradiance in Individual Caesium Lead Halide Quantum Dots. *Nature*
         **2024**, *626*, 535–541.

(45)     Singh, A.; Jana, M. K.; Mitzi, D. B. Reversible Crystal – Glass Transition in a Metal
         Halide Perovskite. *Adv. Mater.* **2020**, *33*, 2005868.

(46)     Kang, J.; Wang, L.-W. High Defect Tolerance in Lead Halide Perovskite $CsPbBr_3$. *J.
         Phys. Chem. Lett.* **2017**, *8*, 489–493.

(47)     Wan, S.; Li, K.; Zou, M.; Hong, D.; Xie, M.; Tan, H.; Scheblykin, I. G.; Tian, Y. All-
         Optical Switching Based on Sub-Bandgap Photoactivation of Charge Trapping in Metal
         Halide Perovskites. *Adv. Mater.* **2023**, *35*, 2209851.

(48)     Marunchenko, A. A.; Baranov, M. A.; Ushakova, E. V.; Ryabov, D. R.; Pushkarev, A.
         P.; Gets, D. S.; Nasibulin, A. G.; Makarov, S. V. Single-Walled Carbon Nanotube Thin
         Film for Flexible and Highly Responsive Perovskite Photodetector. *Adv. Funct. Mater.*
         **2022**, *32*, 2109834.

(49)     Marunchenko, A.; Kondratiev, V.; Pushkarev, A.; Khubezhov, S.; Baranov, M.;
         Nasibulin, A.; Makarov, S. Mixed Ionic-Electronic Conduction Enables Halide-
         Perovskite Electroluminescent Photodetector. *Laser Photonics Rev.* **2023**, *17*, 2300141.

(50)     Wang, H.; Bao, Y.; Li, J.; Li, D.; An, M.; Tang, L.; Li, J.; Tang, H.; Chi, Y.; Xu, J.;
         Yang, Y. Highly Anisotropic Polarization Induced by Electrical Poling in Single-
         Crystalline All-Inorganic Perovskite Nanoplates. *J. Phys. Chem. Lett.* **2023**, *14*, 9943–
         9950.



(51)  Cai, J.; Zhao, T.; Chen, M.; Su, J.; Shen, X.; Liu, Y.; Cao, D. Ion Migration in the All-Inorganic Perovskite CsPbBr$_3$ and Its Impacts on Photodetection. *J. Phys. Chem. C* **2022**, *126*, 10007–10013.

(52)  Ye, Y.-C.; Li, Y.-Q.; Cai, X.-Y.; Zhou, W.; Shen, Y.; Shen, K.-C.; Wang, J.-K.; Gao, X.; Zhidkov, I. S.; Tang, J.-X. Minimizing Optical Energy Losses for Long-Lifetime Perovskite Light-Emitting Diodes. *Adv. Funct. Mater.* **2021**, *31*, 2105813.

(53)  Ma, T.; An, Y.; Yang, Z.; Ai, Z.; Zhang, Y.; Wang, C.; Li, X. Thermodynamic Processes of Perovskite Photovoltaic Devices: Mechanisms, Simulation, and Manipulation. *Adv. Funct. Mater.* **2023**, *33*, 2212596.

(54)  Liao, M.; Shan, B.; Li, M. In Situ Raman Spectroscopic Studies of Thermal Stability of All-Inorganic Cesium Lead Halide (CsPbX$_3$, X = Cl, Br, I) Perovskite Nanocrystals. *J. Phys. Chem. Lett.* **2019**, *10*, 1217–1225.

(55)  Lee, W.; Li, H.; Wong, A. B.; Zhang, D.; Lai, M.; Yu, Y.; Kong, Q.; Lin, E.; Urban, J. J.; Grossman, J. C.; Yang, P. Ultralow Thermal Conductivity in All-Inorganic Halide Perovskites. *Proc. Natl. Acad. Sci. U. S. A* **2017**, *114*, 8693–8697.

(56)  Zhao, Z.; Zhong, M.; Zhou, W.; Peng, Y.; Yin, Y.; Tang, D.; Zou, B. Simultaneous Triplet Exciton−Phonon and Exciton−Photon Photoluminescence in the Individual Weak Confinement CsPbBr$_3$ Micro/Nanowires. *J. Phys. Chem. C* **2019**, *123*, 25349−25358.

(57)  Dresselhaus, M. S.; Dresselhaus, G.; Saito, R.; Jorio, A. Raman Spectroscopy of Carbon Nanotubes. *Phys. Rep.* **2005**, *409*, 47–99.

(58)  Novotny, L.; Hecht, B. *Principles of Nano-Optics*; Cambridge University Press, **2012**.

(59)  Wilmshurst, J. K. Lattice-Type Vibrations in Associated Liquids and the Origin of Anomalous Rayleigh Scattering. *Nature* **1961**, *192*, 1061–1062.

(60)  Protesescu, L.; Yakunin, S.; Bodnarchuk, M. I.; Krieg, F.; Caputo, R.; Hendon, C. H.; Yang, R. X.; Walsh, A.; Kovalenko, M. V. Nanocrystals of Cesium Lead Halide





Perovskites (CsPbX$_3$, X = Cl, Br, and I): Novel Optoelectronic Materials Showing Bright Emission with Wide Color Gamut. *Nano Lett*. **2015**, *15*, 3692–3696.

(61)    Brenner, P.; Bar-On, O.; Jakoby, M.; Allegro, I.; Richards, B. S.; Paetzold, U. W.; Howard, I. A.; Scheuer, J.; Lemmer, U. Continuous Wave Amplified Spontaneous Emission in Phase-Stable Lead Halide Perovskites. *Nat. Commun.* **2019**, *10*, 988.

(62)    Samuel, I. D. W.; Namdas, E. B.; Turnbull, G. A. How to Recognize Lasing. *Nat. Photonics* **2009**, *3*, 546–549.

(63)    Sapienza, R. Controlling Random Lasing Action. *Nat. Phys.* **2022**, *18*, 976–979.

(64)    Khmelevskaia, D.; Markina, D.; Tonkaev, P.; Masharin, M.; Peltek, A.; Talianov, P.; Baranov, M. A.; Nikolaeva, A.; Zyuzin, M. V.; Zelenkov, L. E.; Pushkarev, A. P.; Rogach, A. L.; Makarov, S. V. Excitonic versus Free-Carrier Contributions to the Nonlinearly Excited Photoluminescence in CsPbBr$_3$ Perovskites. *ACS Photonics* **2022**, *9*, 179−189.

(65)    Stranks, S. D.; Burlakov, V. M.; Leijtens, T.; Ball, J. M.; Goriely, A.; Snaith, H. J. Recombination Kinetics in Organic-Inorganic Perovskites: Excitons, Free Charge, and Subgap States. *Phys. Rev. Appl.* **2014**, *2*, 034007.

(66)    Yazdani, N.; Bodnarchuk, M. I.; Bertolotti, F.; Masciocchi, N.; Fureraj, I.; Guzelturk, B.; Cotts, B. L.; Zajac, M.; Rainò, G.; Jansen, M.; Boehme, S. C.; Yarema, M.; Lin, M.-F.; Kozina, M.; Reid, A.; Shen, X.; Weathersby, S.; Wang, X.; Vauthey, E.; Guagliardi, A.; et al. Coupling to Octahedral Tilts in Halide Perovskite Nanocrystals Induces Phonon-Mediated Attractive Interactions Between Excitons. *Nat. Phys.* **2024**, *20*, 47–53.

(67)    Ye, S.; Zhao, M.; Yu, M.; Zhu, M.; Yan, W.; Song, J.; Qu, J. Mechanistic Investigation of Upconversion Photoluminescence in All-Inorganic Perovskite CsPbBrI$_2$ Nanocrystals. *J. Phys. Chem. C* **2018**, *122*, 3152–3156.

(68)    Kajino, Y.; Otake, S.; Yamada, T.; Kojima, K.; Nakamura, T.; Wakamiya, A.;



Kanemitsu, Y.; Yamada, Y. Anti-Stokes Photoluminescence from CsPbBr$_3$ Nanostructures Embedded in a Cs$_4$PbBr$_6$ Crystal. *Phys. Rev. Mater.* **2022**, *6*, L043001.

(69)     Zhang, W.; Ye, Y.; Li, K.; Liu, C. Anti-Stokes Photoluminescence Regulated by Lattice Distortion of CsPbX$_3$ Nanocrystals in Glasses. *J. Phys. Chem. C* **2022**, *126*, 6678–6685.

(70)     Zhou, X.; Zhang, Z. Electron–Phonon Coupling in CsPbBr$_3$. *AIP Adv.* **2020**, *10*, 125015.

(71)     Zhang, M.-L.; Drabold, D. A. Phonon Driven Transport in Amorphous Semiconductors: Transition Probabilities. *Eur. Phys. J. B* **2010**, *77*, 7−23.

(72)     Patel, J. B.; Wright, A. D.; Lohmann, K. B.; Peng, K.; Xia, C. Q.; Ball, J. M.; Noel, N. K.; Crothers, T. W.; Wong-Leung, J.; Snaith, H. J.; Herz, L. M.; Johnston, M. B. Light Absorption and Recycling in Hybrid Metal Halide Perovskite Photovoltaic Devices. *Adv. Mater.* **2020**, *10*, 1903653.

(73)     Sekh, T. V., Cherniukh, I.; Kobiyama, E.; Sheehan, T. J.; Manoli, A.; Zhu, C.; Athanasiou, M.; Sergides, M.; Ortikova, O.; Rossell, M. D.; Bertolotti, F.; Guagliardi, A.; Masciocchi, N.; Erni, R.; Othonos, A.; Itskos, G.; Tisdale, W. A.; Stöferle, T.; Rainò, G.; Bodnarchuk, M. I.; Kovalenko, M. V. All-Perovskite Multicomponent Nanocrystal Superlattices. *ACS Nano* **2024**, *18*, 8423–8436.

(74)     Marunchenko, A.; Kumar, J.; Kiligaridis, A.; Tatarinov, D.; Pushkarev, A.; Vaynzof, Y.; Scheblykin, I. G. Memlumor: A Luminescent Memory Device for Energy-Efficient Photonic Neuromorphic Computing. *ACS Energy Lett.* **2024**, *9*, 2075−2082.